\documentclass[preprint,preprintnumbers,nofootinbib]{revtex4}

\usepackage[dvips]{color}
\usepackage{graphicx}
\usepackage{bbm}

\newcommand{\hif}{\mathchar`-}

\newcommand{\diag}{{\rm Diag}}
\newcommand{\ev}{{\rm eV}}

\newcommand{\gev}{{\rm GeV}}

\newcommand{\bmx}{\left(\begin{array}}
\newcommand{\emx}{\end{array}\right)}
\begin{document}

\title{Non-zero $\theta_{13}$ in models for hierarchical 
neutrino mass spectrum}
\author{
Takeshi Araki\footnote{araki@ihep.ac.cn} } 
\affiliation{
Institute of High Energy Physics, Chinese Academy of
Sciences, Beijing 100049, China }

\begin{abstract}
We introduce three right-handed Majorana neutrinos 
and combine the type-I seesaw and inert doublet 
mechanisms.
The resultant (active) neutrino mass 
matrix is divided into ${\rm rank}=1$ and 
$=2$ parts with different energy scales. 
The different energy scales are reduce to 
different mass scales in the hierarchical 
neutrino mass spectrum.
We apply this scheme to both the inverted and normal 
hierarchy cases and find a correlation between the 
smallest mixing angle ($\theta_{13}$) and the lightest 
neutrino mass.
\end{abstract}

\maketitle

\section{introduction}
Thanks to neutrino 
oscillation experiments \cite{pdg}, we currently have 
convincing evidence that neutrinos have tiny masses and 
mix with each other through the Maki-Nakagawa-Sakata (MNS) 
leptonic mixing matrix.
The recent global analysis of neutrino 
oscillation data yields the following best-fit values
and $1\sigma$ errors \cite{gfit}:
\begin{eqnarray}
&&\Delta m_{21}^2 = 
(7.59 \pm 0.20) \times 10^{-5}\ \ev^2 , \nonumber \\
&&\Delta m_{31}^2 = 
\left\{\begin{array}{l}
-(2.36 \pm 0.11) \times 10^{-3}\ \ev^2\ \ \ {\rm for\ inverted\ hierarchy} \\ 
+(2.46 \pm 0.12) \times 10^{-3}\ \ev^2\ \ \ {\rm for\ normal\ hierarchy} 
\end{array}\right. , \label{eq:exp}\\
&&\theta_{12} = (34.4\pm 1.0)^\circ,\ \ 
\theta_{23} = ( 42.8^{+4.7}_{-2.9} )^\circ,\ \ 
\theta_{13} = ( 5.6^{+3.0}_{-2.7} )^\circ\ ,\nonumber
\end{eqnarray}
which indicate a bi-large mixing pattern 
and leave open three possibilities for the neutrino 
mass spectrum: the normal hierarchy 
($m_3 \gg m_2 > m_1$), inverted hierarchy 
($m_2 > m_1 \gg m_3$) and quasi-degenerate 
($m_1 \simeq m_2 \simeq m_3$) spectra.
The individual neutrino masses as well as the correct 
mass spectrum remain unclear.

On the theoretical side, some extensions of the 
standard model (SM) to accommodate the tiny 
neutrino masses have been proposed.
For instance, in the seesaw mechanisms 
\cite{type1,type2,type3}, new heavy particles are 
introduced to generate neutrino masses suppressed 
by mass scales of the heavy particles, 
while such small neutrino masses can radiatively be 
induced from a loop diagram \cite{zee,zee-babu,idm}, 
too.
Concerning the mixing, 
many constant-number-parametrizations 
(e.g., the democratic \cite{DC}, bi-maximal \cite{BM} 
and tri-bimaximal \cite{TB} mixings) 
have been invented and a lot of efforts have been 
devoted to deriving them from a flavor symmetry.
One of the most attractive features of these 
parametrizations is that they do not depend on the 
neutrino masses, so that no parameter tuning is required 
to obtain the desired mixing pattern.
However, at the same time, it appears that this feature 
have made the mystery of the neutrino mass spectrum 
fade into the background.
Theoretical studies on the mass spectrum seem subtle 
in comparison with those on the mixing: 
we still do not have any plausible model which 
can explain why only $m_3$ stands alone 
whereas $m_1$ and $m_2$ can be nearly degenerate 
in the hierarchical mass spectra, or why they are so 
degenerate in the quasi-degenerate spectrum.

In this Letter, we focus on the 
hierarchical mass spectra and explore 
a possibility that a mass generation mechanism for 
the lighter neutrino(s) is different from that for 
the heavier ones(one). 
Particularly, we consider the following two 
specific scenarios:
\begin{itemize}
\item Scenario-A \\
The mass ordering is inverted. 
At the tree level, $m_{1,2}$ are non-zero and 
completely degenerate, while $m_3$ is vanishing. 
A small $m_3$ and mass splitting between $m_1$ and $m_2$ 
arise from radiative corrections.
\item Scenario-B \\
The mass ordering is normal.
Only $m_3$ is non-zero at the tree level.
$m_1$ and $m_2$ become non-zero after taking 
radiative corrections into account.
\end{itemize}
To this end, we combine the type-I seesaw \cite{type1} 
and inert doublet \cite{idm} mechanisms.
The idea was originally proposed in Ref. \cite{lop-idm} 
to simultaneously explain the relic abundance of 
dark matter, constrains from leptonic processes and 
the baryon asymmetry of the universe as well as 
the neutrino oscillation data.
Here we take a closer look at the neutrino masses 
and try to find possible implications 
for the mixing angles; especially we are 
interested in correlations with the smallest 
mixing angle $\theta_{13}$.
Similar studies are done in Refs. \cite{grimus,type12} 
with a different particle content and/or setup.

This Letter is organized as follows.
In Sec. II, we show a basic framework of 
our scheme and apply it to Scenario-A. 
We investigate Scenario-B in Sec. III and 
summarize our discussion in Sec. IV.

\section{Scenario-A}
\subsection{basic framework}
\begin{table}
\begin{tabular}{|c||c|c|c|c|c|}\hline
           & $L_i$ & $N_S$ 
           & $N_{I}$ & $H$ & $\eta$ \\ \hline
 $SU(2)_L$ & $2$ & $1$ & $1$ 
           & $2$ & $2$ \\ \hline
 $Z_2$     & $+$ & $-$ & $+$ 
           & $+$ & $-$ \\ \hline
\end{tabular}
\caption{
The particle content and charge assignments in Scenario-A.
}
\label{tab:SA}
\end{table}
We extend the SM by introducing three right-handed 
Majorana neutrinos, $N$, and an inert $SU(2)_L$ doublet 
scalar, $\eta$, with a $Z_2$ symmetry.
The particle content and charge assignments 
are summarized in Table \ref{tab:SA}.
We require that all three right-handed neutrinos 
have super-heavy masses, say, 
$M = {\cal O}(10^{10\sim 12})~\gev$, and that 
$\eta$ acquires a zero vacuum-expectation-value (VEV).
The Lagrangian relevant to the following discussions 
is given by
\begin{eqnarray}
{\cal L} =
  Y_{H} ~\overline{L}_i \tilde{H} N_I 
+ Y_{\eta} ~\overline{L}_i \tilde{\eta} N_S 
+ \frac{1}{2}~ M_S N_S N_S
+ \frac{1}{2}~ M_I N_I N_I
+ h.c.~,
\end{eqnarray}
\begin{eqnarray}
V&=&
  \mu_1^2 ~H^\dag H + \mu_2^2 ~\eta^\dag \eta
+ \frac{\lambda_1}{2}(H^\dag H)^2
+ \frac{\lambda_2}{2}(\eta^\dag \eta)^2
+\lambda_3 (H^\dag H)(\eta^\dag \eta) \nonumber \\
&&
+ \lambda_4 (H^\dag \eta)(\eta^\dag H)
+ \frac{\lambda_5}{2}[(H^\dag\eta)^2 + h.c.]~,
\end{eqnarray}
where $L_i$ stands for the left-handed $SU(2)_L$ 
doublet leptons and $H$ denotes the SM Higgs field 
with $\tilde{H}=i\sigma_2 H^*$.
The subscript $i$ runs over 
$1$ to $3$ while $I$ is $1$ or $2$.
Thus, $Y_H$ and $Y_\eta$ are $3\times 2$ and $3\times 1$ 
dimensional matrices, respectively.
Notice that we have chosen the basis in which the 
charged lepton and right-handed neutrino 
mass matrices are diagonal, real and positive, 
and the real basis of $\lambda_5$ and $Y_\eta$.

By implementing the type-I seesaw mechanism, 
we obtain the following 
tree-level neutrino mass matrix:
\begin{eqnarray}
M^0 
=\frac{v^2}{M_1}
\bmx{ccc}
A^2 & AB & AC \\
AB & B^2 & BC \\
AC & BC & C^2
\emx
+\frac{v^2}{M_2}
\bmx{ccc}
D^2 & DE & DF \\
DE & E^2 & EF \\
DF & EF & F^2
\emx~, \label{eq:AM0}
\end{eqnarray}
where $v=174~\gev$ is the VEV of the SM Higgs field and 
$A\cdots F$ are complex Yukawa couplings included in $Y_H$.
Besides, we can induce a one-loop neutrino mass operator 
by exchanging $N_S$ and $\eta^0$ \cite{idm}, 
and it results in 
\begin{eqnarray}
\delta M
=
\frac{v^2}{M_S}
\bmx{ccc}
\alpha^2 & \alpha\beta & \alpha\gamma \\
\alpha\beta & \beta^2 & \beta\gamma \\
\alpha\gamma & \beta\gamma & \gamma^2
\emx 
\frac{\lambda_5}{8\pi^2}
\left[ \ln\frac{M_S^2}{m_\eta^2}-1 \right]~,
\label{eq:AdM}
\end{eqnarray}
where $\alpha\cdots \gamma$ are real Yukawa couplings 
included in $Y_\eta$. 
In Eq. (\ref{eq:AdM}), we have defined 
$m_\eta^2 \equiv \mu_2^2 + (\lambda_3 + \lambda_4)v^2$ 
and assumed $M_S^2 \gg m_\eta^2 \gg 2\lambda_5 v^2$ 
for simplicity.
As one can see from Eqs. (\ref{eq:AM0}) and 
(\ref{eq:AdM}), the tree-level ($M^0$) and 
one-loop ($\delta M$) mass matrices 
are ${\rm rank}=2$ and $1$, respectively, 
with different energy scales.
Since $\delta M$ is suppressed with 
$\lambda_5/8\pi^2$ in comparison with $M^0$ in 
the case of $M_S \simeq M_I$, we conjecture that 
$M^0$ is responsible for the heavier-neutrino masses 
($m_{1,2}$) and the lightest neutrino mass ($m_3$) 
originates in $\delta M$.
Thus, this scheme suggests the inverted hierarchy 
spectrum.

\subsection{neutrino masses and mixing}
We apply the above scheme to Scenario-A and look at 
the neutrino mixing.
Let us suppose that there exists a low-energy\footnote{
We ignore corrections due to the RGE running effects.
} 
flavor symmetry which guarantees $\theta_{13}=0^\circ$ 
at the tree level.
Hence, we consider the following tree-level mixing matrix:
\begin{eqnarray}
V^0 = 
\left(\begin{array}{ccc}
 c_{12}^0 & s_{12}^0 & 0 \\
 -s_{12}^0 c_{23}^0 & c_{12}^0 c_{23}^0 & s_{23}^0 \\
 s_{12}^0 s_{23}^0 & -c_{12}^0 s_{23}^0 & c_{23}^0
\end{array}\right)~, \label{eq:V0}
\end{eqnarray}
where 
$c_{ij}^0(s_{ij}^0)=\cos\theta_{ij}^0(\sin\theta_{ij}^0)$.
However, once we insist the degeneracy between $m_1$ 
and $m_2$ at the tree level, $M^0$ in Eq. (\ref{eq:AM0}) 
may take the form of 
\begin{eqnarray}
M^0 = V^0~\diag(m_0,m_0,0)~(V^0)^T = m_0
\bmx{ccc}
1 & 0 & 0 \\
0 & (c_{23}^0)^2 & -s_{23}^0 c_{23}^0 \\
0 & -s_{23}^0 c_{23}^0 & (s_{23}^0)^2
\emx \label{eq:AM02}
\end{eqnarray}
with a complex parameter $m_0$, 
and this mass matrix is diagonalized by 
only $\theta_{23}$.
Thus, we start the discussion with Eqs. (\ref{eq:AM02}) 
and (\ref{eq:V0})  with $\theta_{12}^0=0^\circ$ 
at the tree level.
Non-zero $\theta_{12}$, $\theta_{13}$, $m_3$ and the 
mass splitting between $m_1$ and $m_2$ will arise after 
diagonalizing the full mass matrix 
$M_\nu = M^0 + \delta M$ with the full mixing matrix
\begin{eqnarray}
V=
V^0
\left(\begin{array}{ccc}
 1 & 0 & 0 \\
 0 &  c_{23}^d & s_{23}^d \\
 0 & -s_{23}^d & c_{23}^d
\end{array}\right)
\left(\begin{array}{ccc}
 c_{13} & 0 & s_{13}\ e^{-i\delta} \\
 0 &  1 & 0 \\
 -s_{13}\ e^{i\delta} & 0 & c_{13}
\end{array}\right)
\left(\begin{array}{ccc}
  c_{12} & s_{12} & 0 \\
 -s_{12} &  c_{12} & 0 \\
 0 & 0 & 1
\end{array}\right)\Omega,
\end{eqnarray}
where 
$\theta_{23}^d=\theta_{23}- \theta_{23}^0$ and
$\Omega$ contains two Majorana CP-violating phases. 

Prior to showing the results of numerical calculations, 
it may be useful to derive some approximate expressions 
of the mixing angles and masses.
By taking the limit of $(s_{23}^d)^2 = 0$, 
we arrive at
\begin{eqnarray}
&&\tan 2\theta_{12}\simeq
\frac{ 2\alpha (\beta c_{23}-\gamma s_{23})c_{13}~\delta m}
{(c_{13}^2-1)m_0+(\alpha c_{13})^2 \delta m - (\beta c_{23}-\gamma s_{23})^2 \delta m}~,  \\
&&\tan 2\theta_{13}\simeq
\frac{2\alpha(\beta s_{23}+\gamma c_{23})\delta m}
{\left|(m_0+\alpha^2 \delta m)e^{i\delta}
-(\beta s_{23}+\gamma c_{23})^2 \delta m~e^{-i\delta}\right|} 
~, \\
&&m_3 \simeq 
\left[
(\beta s_{23}+\gamma c_{23})c_{13}
\right]^2 \delta m ~,
\end{eqnarray}
where
\begin{eqnarray}
\delta m = 
\frac{v^2}{M_S}
\frac{\lambda_5}{8\pi^2}
\left[ \ln\frac{M_S^2}{m_\eta^2}-1 \right]
\label{eq:dm}
\end{eqnarray}
and $\alpha$, $\beta$ and $\gamma$ are real 
Yukawa couplings defined in Eq. (\ref{eq:AdM}).
Notice that we have omitted some terms associated 
with $s_{13}$ in the expressions of 
$\tan2\theta_{12}$ and $m_3$.
From the above expressions, one can see that 
$\theta_{12}$, $\theta_{13}$ and $m_3$ are not 
sensitive to the initial value of $\theta_{23}$
and find interesting correlations among them: 
e.g., when $\theta_{13}$ is non-zero (or zero), $m_3$ 
is also non-zero (or zero) since $\theta_{12}\neq 0$ 
restricts $\alpha$ to be non-zero.
\begin{figure}[t]
\begin{center}
\includegraphics*[width=0.48\textwidth]{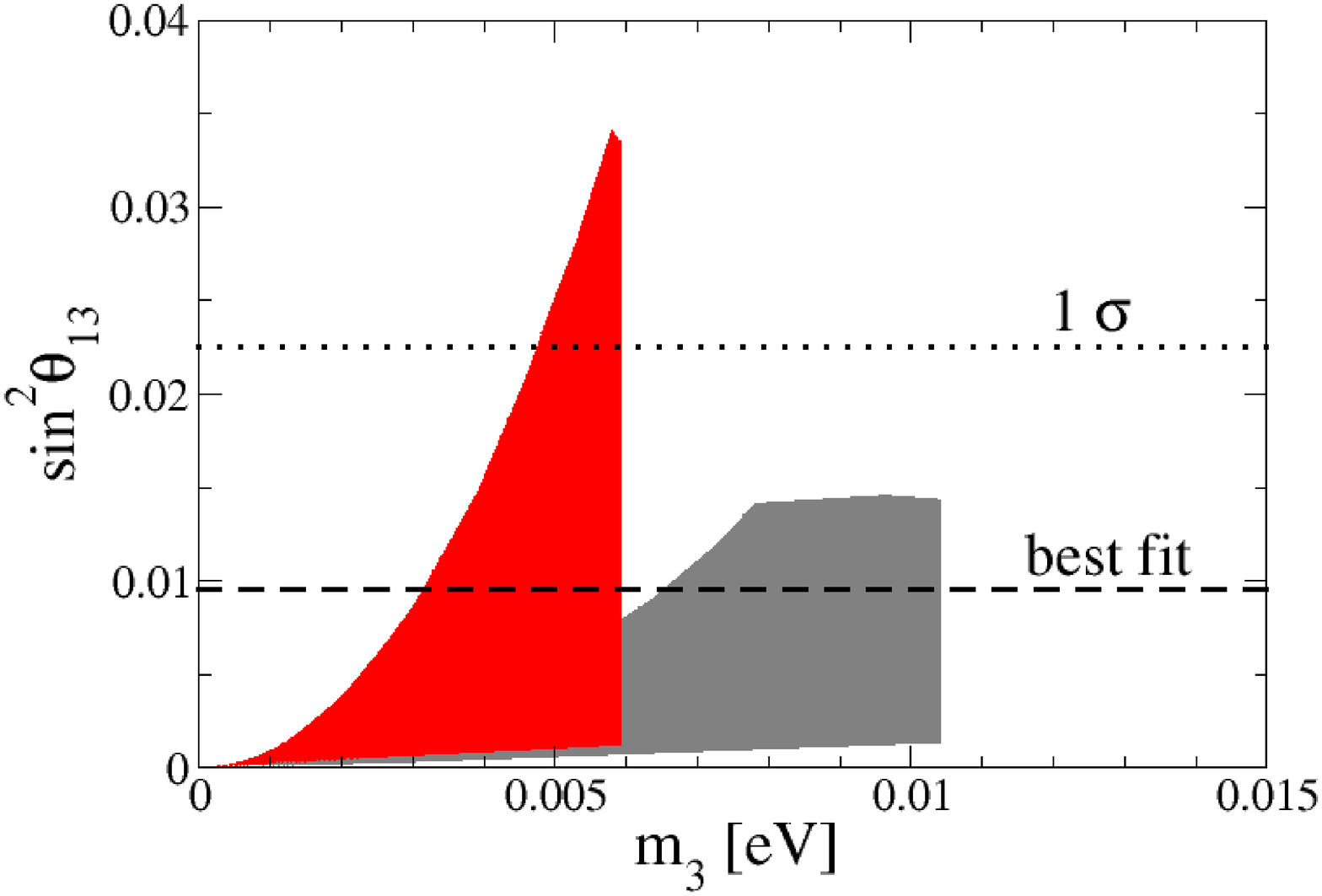}
\includegraphics*[width=0.48\textwidth]{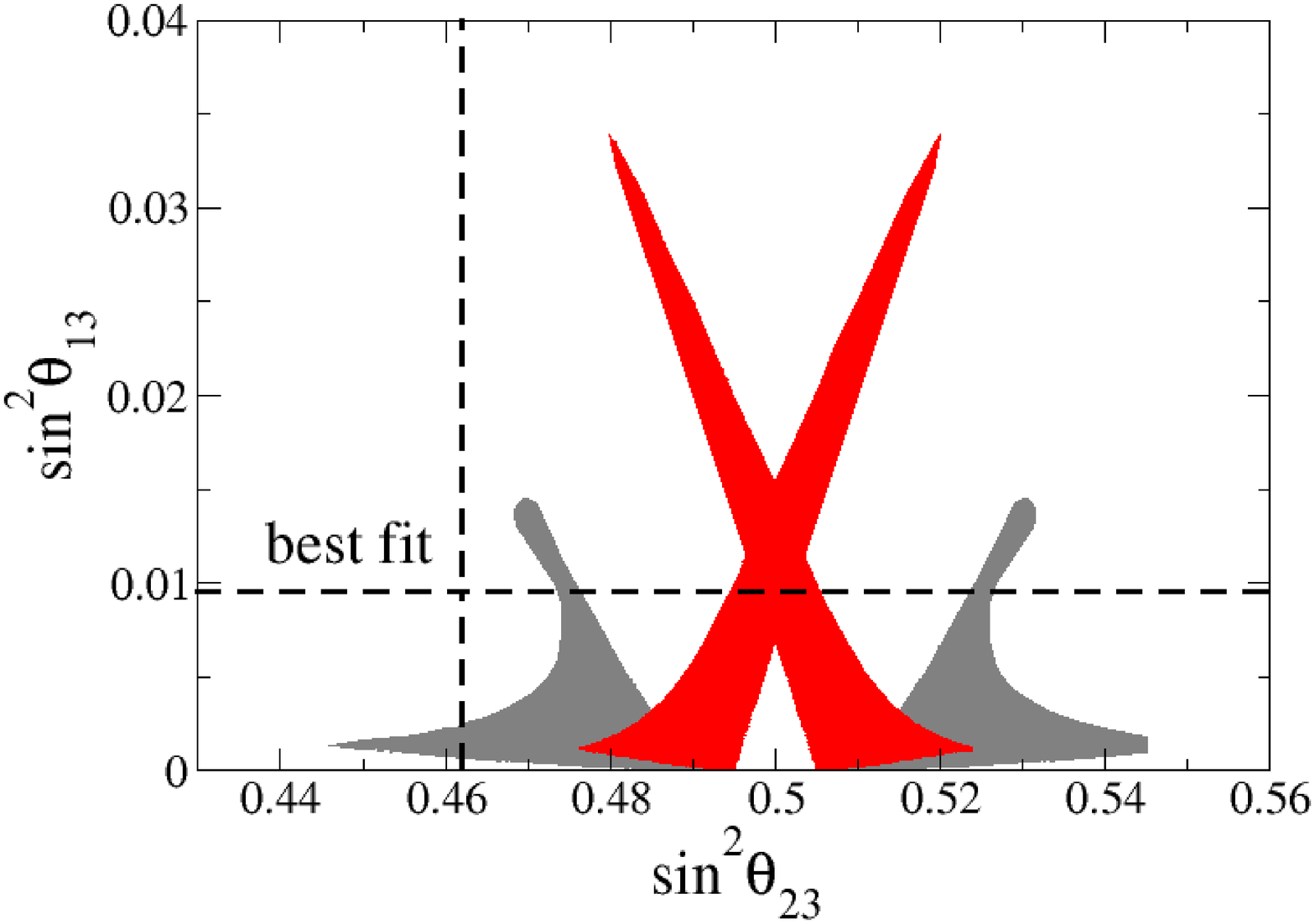}
\caption{\footnotesize 
$\sin^2\theta_{13}$ as a function of the lightest 
neutrino mass, $m_3$, (left panel) 
and $\sin^2\theta_{23}$ (right panel) in Scenario-A.
In the red (gray) region, $m_1$ ($m_2$) is slightly 
perturbed and decreased (increased) while corrections 
for $m_2$ ($m_1$) are negligibly small. 
The dotted and dashed lines display the $1\sigma$ 
upper bound of $\sin^2\theta_{13}$ and best-fit 
values of $\sin^2\theta_{13}$ and $\sin^2\theta_{23}$, 
respectively.
} \label{fig:SA}
\end{center}
\end{figure}
This correlation is not the result of the 
approximation we made.
In Fig. \ref{fig:SA}, we numerically diagonalize 
the full neutrino mass matrix in the case of 
$\theta_{23}^0=45^\circ$ and plot 
$\sin^2\theta_{13}$ as a function of the lightest 
neutrino mass, $m_3$, (left panel) with respect to
the $1\sigma$ constraints of $\Delta m^2_{21}$, 
$\Delta m^2_{31}$, $\theta_{12}$ and $\theta_{23}$ 
given in Eq. (\ref{eq:exp}).
Since we are focusing on the hierarchical neutrino 
mass spectrum, we have fixed the absolute value of 
$m_0$ by the best-fit value of $\Delta m^2_{31}$, 
i.e., $|m_0|=\sqrt{2.36\times 10^{-3}}~~\ev$, 
while varying its phase within $0$ to $360^\circ$. 
As can be seen from the figure, there are two 
parameter regions in this model: 
in the red (gray) region, $m_1$ ($m_2$) is slightly 
perturbed and decreased (increased) while corrections 
for $m_2$ ($m_1$) are negligibly small. 
Nevertheless, the corrections for $m_1$ are sufficiently 
small in comparison with $|m_0|$ and thus, $m_1$ 
can approximately be given by $m_1 \simeq |m_0|$.
Therefore, the $1\sigma$ constraint of 
$\Delta m^2_{31}$ can be translated into an upper 
bound on $m_3$, which places an upper bound on 
$\theta_{13}$ and one can read off 
$\sin^2\theta_{13}<0.034$ ($\theta_{13} < 10.6^\circ$) 
from the red region. 
Interestingly, this upper bound is consistent with 
the recently reported T2K and MINOS results 
\cite{t2k,MINOS}, which indicate a relatively large 
$\theta_{13}$.

We also plot $\sin^2\theta_{13}$ as a function of 
$\sin^2\theta_{23}$ in the right panel. 
In the red regions, $\sin^2\theta_{23}$ stays within 
$0.50 \pm 0.02$, while it can largely 
deviate from the initial value in the gray regions.

We remark that corrections to $\theta_{12}$ can 
in general be enhanced by the near degeneracy between 
$m_1$ and $m_2$ \cite{agx}.
Therefore, we can always account for 
$\theta_{12}\simeq 34^\circ$ 
even starting from $0^\circ$.

\subsection{CP violation}
\begin{figure}[t]
\begin{center}
\includegraphics*[width=0.6\textwidth]{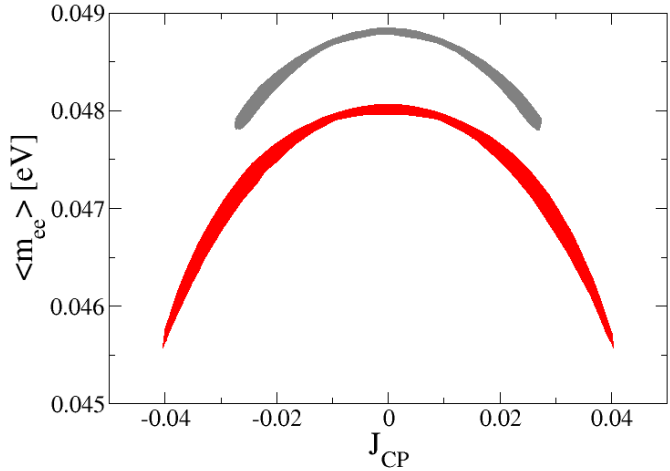}
\caption{\footnotesize
The rephasing-invariant Jarlskog parameter, $J_{CP}$, 
and the effective mass, $\langle m_{ee}\rangle$, of 
neutrinoless double beta decay.
The legend of colored regions is the same as 
Fig. \ref{fig:SA}.
} \label{fig:CP}
\end{center}
\end{figure}
Since $\theta_{13}$ becomes non-zero after taking 
the radiative corrections into account and the model 
is described by a single CP-violating phase, it may be 
interesting to see a correlation between 
the rephasing-invariant Jarlskog parameter:
\begin{eqnarray}
J_{CP}=
{\rm Im}[V_{e2}^{}V_{\mu 3}^{}V_{e3}^{*}V_{\mu 2}^{*}]
\end{eqnarray}
and the effective mass of neutrinoless double beta decay 
($0\nu\beta\beta$):
\begin{eqnarray}
\langle m_{ee}\rangle=
\left| V^2_{e1}m_1 + V^2_{e2}m_2 + V^2_{e3}m_3 \right| .
\end{eqnarray}
In Fig. \ref{fig:CP}, we plot $\langle m_{ee}\rangle$ 
as a function of $J_{CP}$ under the same 
conditions as Fig. \ref{fig:SA}.
We find that the magnitude of $\langle m_{ee}\rangle$ 
is around $0.046 \sim 0.049$, which could be reachable 
in the near future experiments \cite{ovbb}.
Moreover, $J_{CP}$ is expected to be measured at 
long baseline neutrino oscillation experiments. 
Since $J_{CP}$ and $\langle m_{ee}\rangle$ 
are strongly correlated with each other in this 
model, these upcoming experiments may enable us 
to confirm or rule out the model.

\section{Scenario-B}
If we interchange the $Z_2$ assignments of $N_S$ 
and $N_I$ in Table \ref{tab:SA}, the scheme proposed 
in Sec. II-A becomes applicable to the normal 
hierarchy case\footnote{
Alternatively, one can simply assume 
$\delta M \gg M^0$.
}.
In this case, $N_S$ couples to the SM Higgs ($H$) 
while $N_I$ to the inert double ($\eta$). 
Consequently, the tree-level and one-loop mass 
matrices turn out to be 
\begin{eqnarray}
&&M^0 
= \frac{v^2}{M_S}
\bmx{ccc}
\alpha^2 & \alpha\beta & \alpha\gamma \\
\alpha\beta & \beta^2 & \beta\gamma \\
\alpha\gamma & \beta\gamma & \gamma^2
\emx ~, \label{eq:BM0} \\
&&\delta M
= \delta m_1
\bmx{ccc}
A^2 & AB & AC \\
AB & B^2 & BC \\
AC & BC & C^2
\emx
+ \delta m_2
\bmx{ccc}
D^2 & DE & DF \\
DE & E^2 & EF \\
DF & EF & F^2
\emx~, \label{eq:BdM}
\end{eqnarray}
respectively, where the definitions of $\delta m_1$ 
and $\delta m_2$ are similar to that given 
in Eq. (\ref{eq:dm}).

Let us apply this scheme to Scenario-B, namely, 
we presume that $M^0$ is responsible for the heaviest 
neutrino mass ($m_3$) and $\delta M$ for the lighter 
neutrino masses ($m_{1,2}$). 
Also, we employ $V^0$ in Eq. (\ref{eq:V0}) as the 
tree-level mixing matrix. 
As a result, $M^0$ may take the form of
\begin{eqnarray}
M^0 = m_0
\bmx{ccc}
0 & 0 & 0 \\
0 & (s_{23}^0)^2 & s_{23}^0 c_{23}^0 \\
0 & s_{23}^0 c_{23}^0 & (c_{23}^0)^2
\emx \label{eq:BM02}
\end{eqnarray}
and this mass matrix is again diagonalized by 
only $\theta_{23}$. 
The other neutrino masses and mixing angles 
are obtained after including $\delta M$ 
in Eq. (\ref{eq:BdM}).
However, because $\delta M$ contains a lot of 
parameters, we cannot establish correlations 
among the neutrino masses and mixing angles. 
In order to do that, we simplify 
the mass matrix by imposing $C=B$, $F=-E$, 
$\theta_{23}^0=45^\circ$ \footnote{
A discrete flavor symmetry may realize these conditions.
We show a simple realization based on the $D_4$ 
symmetry in Appendix.
} and CP invariance.
Then, the full neutrino mass matrix is given by 
\begin{eqnarray}
M^\prime_\nu
=
\bmx{ccc}
0 & 0 & 0 \\
0 & 0 & 0 \\
0 & 0 & m_0
\emx
+ \delta m_1
\bmx{ccc}
A^2 & 0 & \sqrt{2}AB \\
0 & 0 & 0 \\
\sqrt{2}AB & 0 & 2B^2
\emx
+ \delta m_2
\bmx{ccc}
D^2 & \sqrt{2}DE & 0 \\
\sqrt{2}DE & 2E^2 & 0 \\
0 & 0 & 0
\emx \label{eq:BMn}
\end{eqnarray}
in the diagonal basis of $M^0$.
Roughly speaking, the second and third terms 
originate non-zero $\theta_{13}$ and 
$\theta_{12}$, respectively, and they are 
approximately expressed as 
\begin{eqnarray}
&&\tan2\theta_{12}\simeq
\frac{2\sqrt{2}DE c_{13} ~\delta m_2}
{(2E^2-D^2)\delta m_2 - A^2 \delta m_1}~, \\
&&\tan2\theta_{13}\simeq
\frac{2\sqrt{2}AB\delta m_1}
{(2B^2 -A^2)\delta m_1 - D^2\delta m_2 + m_0}~,
\end{eqnarray}
while corrections for $\theta_{23}$ are negligibly small.
Moreover, $m_1$ and $m_2$ are given by
\begin{eqnarray}
&&m_1 \simeq A^2 c_{12}^2 \delta m_1 
+ (D c_{13}c_{12}-\sqrt{2}E s_{12})^2 \delta m_2 ~, \\
&&m_2 \simeq A^2 s_{12}^2 \delta m_1 
+ (D c_{13}s_{12}+\sqrt{2}E c_{12})^2 \delta m_2 ~.
\end{eqnarray}
\begin{figure}[t]
\begin{center}
\includegraphics*[width=0.48\textwidth]{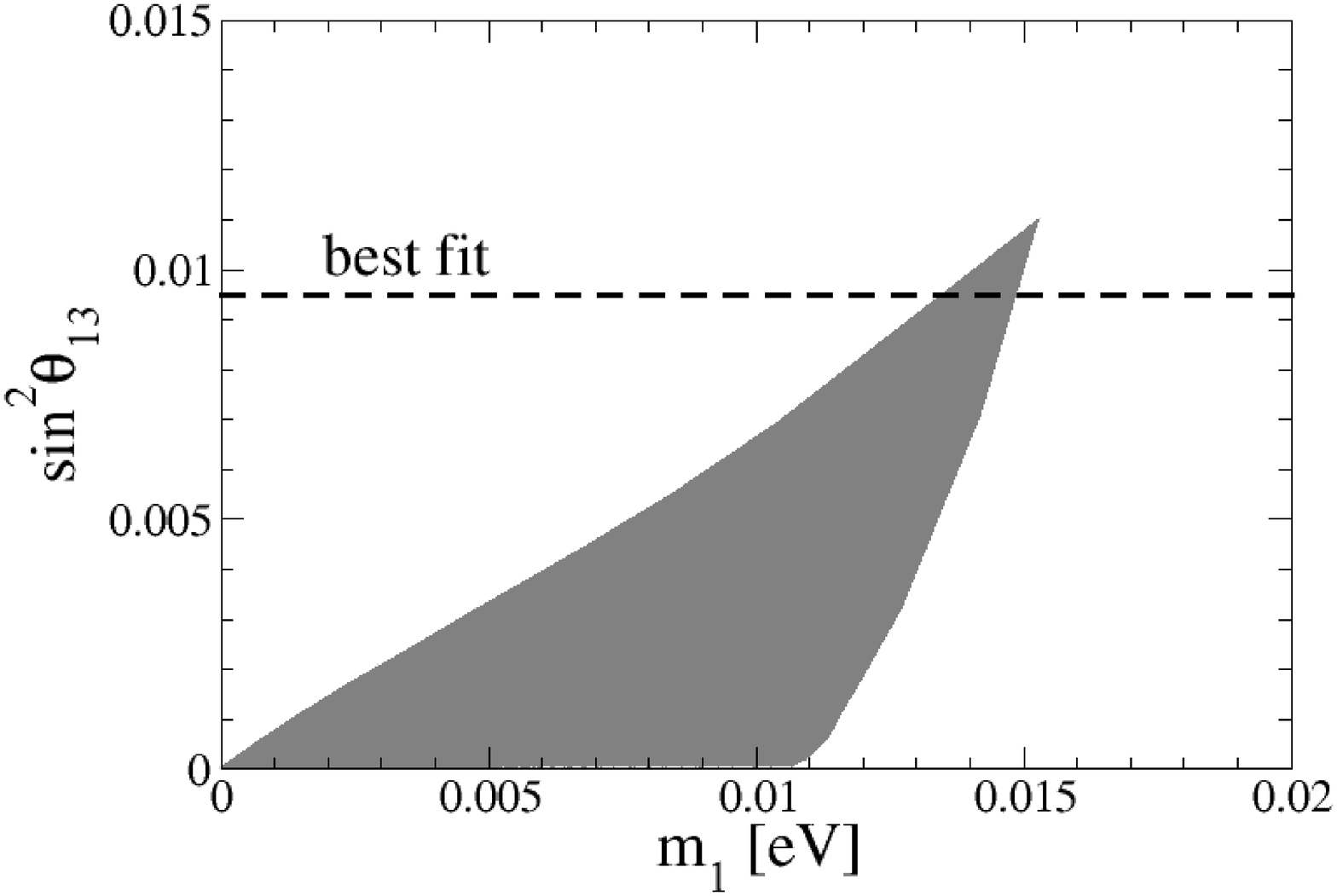}
\includegraphics*[width=0.48\textwidth]{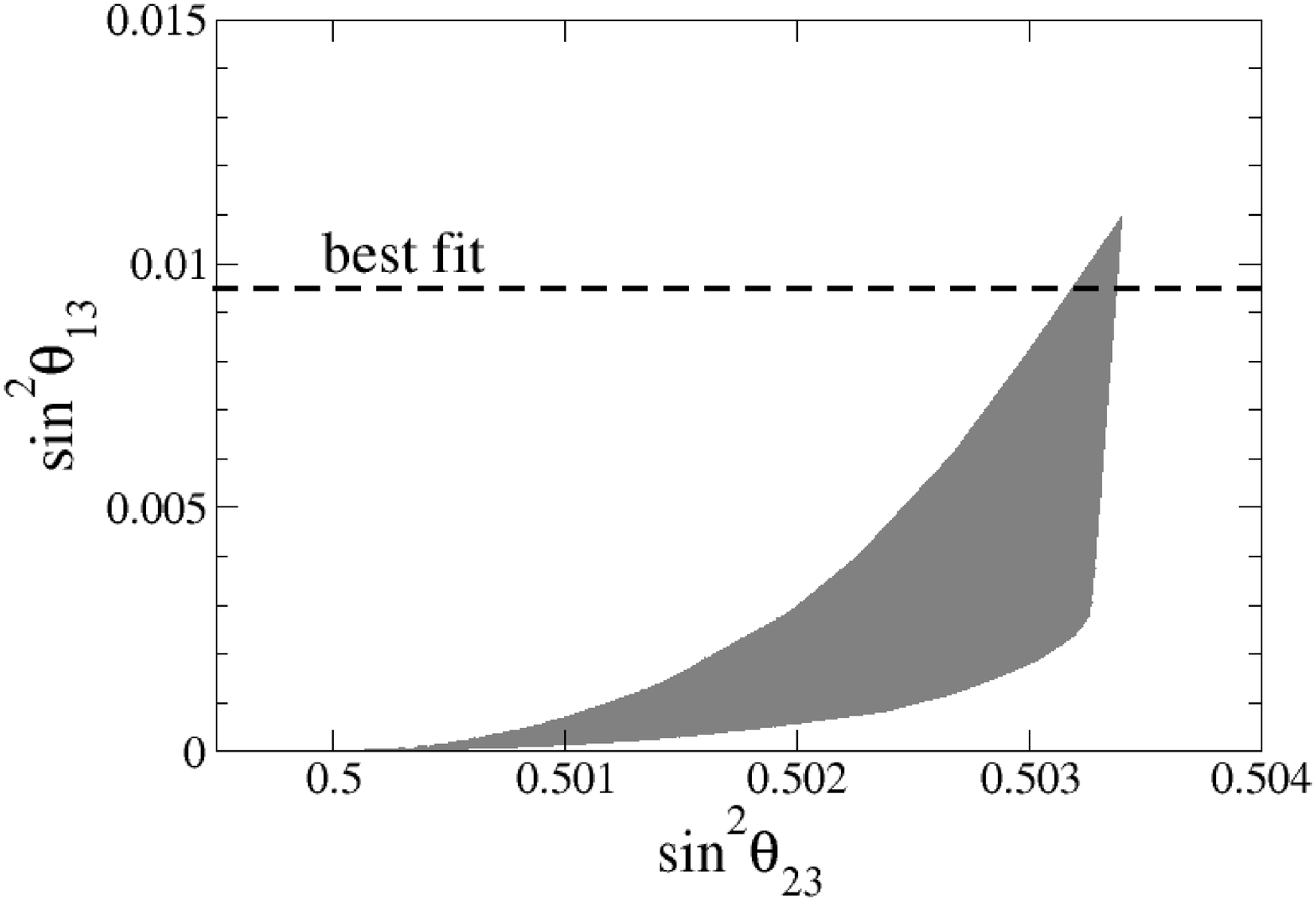}
\caption{\footnotesize 
$\sin^2\theta_{13}$ as a function of the lightest 
neutrino mass, $m_1$, (left panel) 
and $\sin^2\theta_{23}$ (right panel) in Scenario-B.
The dashed lines display the best-fit value 
of $\theta_{13}$.
} \label{fig:SB}
\end{center}
\end{figure}
By requiring 
$m_0 = \sqrt{2.46\times 10^{-3}}~\ev$ and
$1\sigma$ constraints of $\Delta m^2_{21}$, 
$\Delta m^2_{31}$, $\theta_{12}$ and $\theta_{23}$, 
we plot $\sin^2\theta_{13}$ as a function of the 
lightest neutrino mass, $m_1$, (left panel) and 
$\sin^2\theta_{23}$ (right panel) in Fig. \ref{fig:SB}.
We note that corrections to $m_3$ are not 
negligible in this model, so that we have imposed 
$m_3 < \sqrt{(2.46+0.12)\times 10^{-3}}~\ev$ 
in order to keep the hierarchical mass spectrum.
In this case, the $1\sigma$ constraint of 
$\Delta m^2_{31}$ can be translated into an upper 
bound on $m_1$ and it leads to 
$\sin^2\theta_{13}<0.011$ ($\theta_{13}<6.0^\circ$).
Furthermore, $\theta_{23}$ remains almost maximal 
and this model indicates $\theta_{23}>45^\circ$.

\section{Conclusion}
We have considered a combination of the type-I seesaw 
and inert doublet mechanisms with three right-handed 
Majorana neutrinos.
The resultant (active) neutrino mass matrix is divided 
into ${\rm rank}=1$ and $=2$ parts with 
different energy scales, and it suggests the hierarchical 
neutrino mass spectrum.
We have applied this scheme to two 
scenarios in which both the lightest neutrino 
mass and a non-zero $\theta_{13}$ are radiatively 
induced via the inert doublet mechanism. 
We have found that the constraint of $\Delta m_{31}^2$ 
leads to an upper bound for the lightest neutrino mass, 
and it subsequently constraints the size of $\theta_{13}$.
Given the $1\sigma$ constrains of Eq. (\ref{eq:exp}), 
we have obtained $\sin^2\theta_{13}<0.034$ 
($\theta_{13}<10.6^\circ$) in Scenario-A.
In Scenario-B, we have assumed a simple mass texture 
and gained $\sin^2\theta_{13}<0.011$ ($\theta_{13}<6.0^\circ$).

As discussed in Refs. \cite{lop-idm,kms}, this 
kind of scheme possesses a great possibility 
for understanding other phenomena, such as the relic 
abundance of dark matter, some leptonic processes 
and the baryon asymmetry of the universe.
Especially, since we have a unique CP-violating 
phase in Scenario-A, we may be able to directly 
relate the low-energy CP violation with leptogenesis.
Further extensive studies including them could 
make a difference between our scheme and others.
We shall study this issue elsewhere.

\begin{acknowledgments}
I would like to thank J. Kubo and D. Suematsu
for letting me know their previous studies and 
their kind hospitality at Kanazawa university.
I am also grateful to Z.Z. Xing for useful discussions.
This work  was supported in part by the National
Natural Science Foundation of China under Grant 
No. 10875131 and by the Chinese Academy of Sciences 
Fellowship for Young International Scientists. 
\end{acknowledgments}

\appendix\section{$D_4$ flavor model}
\begin{table}
\begin{tabular}{|c||c|c|c|c|c|c|c|c|c|c|}\hline
       & $L_1$ & $L_{D=2,3}$ & $N_S$ & $N_1$ & $N_2$ 
       & $H$ & $\eta$ & $D$ & $S^{''}$ & $S^{'''}$ \\ \hline
 $D_4$ & $1^{'}$ & $2$ & $1$ & $1^{'''}$ & $1^{''}$ 
       & $1$ & $1$ & $2$ & $1^{''}$ & $1^{'''}$ \\ \hline
 $Z_2^\prime$ & $+$ & $+$ & $-$ & $-$ & $-$
              & $+$ & $+$ & $-$ & $-$ & $-$ \\ \hline
\end{tabular}
\caption{
The particle content and charge assignments 
of the $D_4$ model.
}
\label{tab:SB}
\end{table}
We show a simple realization of the mass matrix 
Eq. (\ref{eq:BMn}).
In addition to the $Z_2$ symmetry, we introduce 
$D_4$-flavor and $Z_2^\prime$-auxiliary symmetries 
with gauge singlet 
flavon fields $D$, $S^{''}$ and $S^{'''}$.
The particle content and charged assignments are 
summarized in Table \ref{tab:SB}, and the tensor 
products of $D_4$ are given by \cite{babu-kubo}
\begin{eqnarray}
&&
\begin{array}{ccccccccc}
\bmx{c} x_1 \\ x_2 \emx & \otimes & 
\bmx{c} y_1 \\ y_2 \emx & = & 
(x_1y_1 + x_2y_2) & \oplus & (x_1y_1 - x_2y_2) & & \\
2 & \otimes & 2 & = & 1 & \oplus & 1^{''} & & \\
 & & & & & & & & \\
 & & & & &
\oplus & (x_1y_2 - x_2y_1) & \oplus & (x_1y_2 + x_2y_1)~, \\
 & & & & & \oplus & 
1^{'} & \oplus & 1^{'''}
\end{array} \\ \nonumber \\
&&
1^{'} \otimes 1^{'} = 1^{''} \otimes 1^{''}
= 1^{'''} \otimes 1^{'''} = 1~, \\
&&
1^{'} \otimes 1^{''} = 1^{'''},~~
1^{''} \otimes 1^{'''} = 1^{'},~~
1^{'} \otimes 1^{'''} = 1^{''}~.
\end{eqnarray}
Because of the symmetries, the Lagrangian of the 
neutrino sector is written as 
\begin{eqnarray}
{\cal L}=
&&\frac{\beta}{\Lambda}~\overline{L}_D \tilde{H} N_S D 
+ \frac{A}{\Lambda}~\overline{L}_1 \tilde{\eta} N_1 S^{''} 
+ \frac{B}{\Lambda}~\overline{L}_D \tilde{\eta} N_1 D
\nonumber \\
&&
+ \frac{D}{\Lambda}~\overline{L}_1 \tilde{\eta} N_2 S^{'''} 
+ \frac{E}{\Lambda}~\overline{L}_D \tilde{\eta} N_2 D 
+ {\cal O}(1/\Lambda^3) + \cdots
\nonumber \\
&&
+ \frac{1}{2}~ M_S N_S N_S
+ \frac{1}{2}~ M_1 N_1 N_1
+ \frac{1}{2}~ M_2 N_2 N_2
+ {\cal O}(1/\Lambda) + \cdots + h.c. ~,
\end{eqnarray}
where we have written down only the leading terms 
and $\Lambda$ denotes a typical energy scale of 
the $D_4$ flavor symmetry.
If we demand the VEV alignment: 
$\langle D \rangle \propto (1,1)$, the tree-level 
and one-loop neutrino mass matrices turn out to be
\begin{eqnarray}
&&M^0 
= \frac{v^2}{M_S}
\bmx{ccc}
0 & 0 & 0 \\
0 & \beta^2 & \beta^2 \\
0 & \beta^2 & \beta^2
\emx ~,  \\
&&\delta M
= \delta m_1
\bmx{ccc}
A^2 & AB & AB \\
AB & B^2 & B^2 \\
AB & B^2 & B^2
\emx
+ \delta m_2
\bmx{ccc}
D^2 & DE & -DE \\
DE & E^2 & -E^2 \\
-DE & -E^2 & E^2
\emx~, 
\end{eqnarray}
respectively, where VEVs of the flavons and 
$\Lambda$ are included in the Yukawa couplings.
$M^0$ can be diagonalized by the $45^\circ$ 
rotation in the $2\hif 3$ plane and then, we obtain 
the neutrino mass matrix given in Eq. (\ref{eq:BMn}).
Furthermore, by adding extra Higgs doublets to the 
charged lepton sector, we can easily derive 
a diagonal charged lepton mass matrix \cite{GriLavo}.

\end{document}